\documentclass{osa-article}

\journal{oe}

\articletype{Research Article}

\begin{document}

\title{Quantum Simulation of Tunable and Ultrastrong Mixed-Optomechanics}

\author{Yue-Hui Zhou, Xian-Li Yin, and Jie-Qiao Liao\authormark{*}}

\address{Key Laboratory of Low-Dimensional Quantum Structures and Quantum Control of
Ministry of Education, Key Laboratory for Matter Microstructure and Function of Hunan Province, Department of Physics and Synergetic Innovation
Center for Quantum Effects and Applications, Hunan Normal University,
Changsha 410081, China}

\email{\authormark{*}jqliao@hunnu.edu.cn} 

\begin{abstract}
We propose a reliable scheme to simulate tunable and ultrastrong mixed (first-order and quadratic optomechanical couplings coexisting) optomechanical interactions in a coupled two-mode bosonic system, in which the two modes are coupled by a cross-Kerr interaction and one of the two modes is driven through both the single- and two-excitation processes. We show that the mixed-optomechanical interactions can enter the single-photon strong-coupling and even ultrastrong-coupling regimes. The strengths of both the first-order and quadratic optomechanical couplings can be controlled on demand, and hence first-order, quadratic, and mixed optomechanical models can be realized. In particular, the thermal noise of the driven mode can be suppressed totally by introducing a proper squeezed vacuum bath. We also study how to generate the superposition of coherent squeezed state and vacuum state based on the simulated interactions. The quantum coherence effect in the generated states is characterized by calculating the Wigner function in both the closed- and open-system cases. This work will pave the way to the observation and application of ultrastrong optomechanical effects in quantum simulators.
\end{abstract}

\section{Introduction}

In recent years, cavity optomechanics~\cite{Kippenberg2008Science,Aspelmeyer2012PhysTod,Aspelmeyer2014RMP} has attracted much attention from the communities of quantum physics, quantum optics, and quantum information sciences. This is because cavity optomechanics has significance in both the study of the fundamental problems of quantum mechanics and the quantum precision measurements. So far, much effort has been devoted to both the studies of optomechanical interactions and the applications of optomechanical effects in modern quantum technologies, and great advances have been achieved in this field. These advances include ground-state cooling of moving mirrors~\cite{Wilson-Rae2007PRL,Marquardt2007PRL,Genes2008PRA,Chan2011Nature,TeufelD2011Nature,Lai2018PRA,Lai2020PRAR}, optomechanical entanglement~\cite{Bose1997PRA,Vitali2007PRL,Hartmann2008PRL,Tian2013PRL,Wang2013PRL,Palomaki2013Science,Riedinger2018Nature,Ockeloen2018Nature}, precision detection of weak forces~\cite{Vitali2001PRA,Tsang2010PRL,Clerk2010RMP,Wimmer2014PRA,Motazedifard2016NJP,Zhang2017NJP,Armata2017PRA,Li2018PRA,Zhao2020}, optomechanically induced transparency~\cite{Agarwal2010PRA,Weis2010Science,Safavinaeini2011Nature}, photon blockade effect~\cite{Liao2013PRA,Rabl2011PRL,Xu2013PRA,Liao2013PRAc}, and Kerr nonlinearity~\cite{Aldana2013PRA}. Physically, all these effects are mainly induced by two typical optomechanical interactions: the first-order~\cite{Law1995PRA,Aspelmeyer2014RMP} and quadratic optomechanical~\cite{Thompson2008Nature,Bhattacharya2008PRA,Bhattacharya2008PRAR,Jayich2008NJP,Sankey2010NP,Shi2013PRA,Liao2013PRA,Liao2014SR} couplings, which are proportional to the displacement and the square of the displacement of the mechanical oscillator, respectively. Currently, the strong linearized optomechanical coupling between optical and mechanical modes has been observed in experiments~\cite{Dobrindt2008PRL,Groblacherl2009Nature,Teufel2011Nature,Verhagen2012Nature,Bothner2020Naturephysics}. In particular, the ultrastrong parametric coupling between a superconducting cavity and a mechanical resonator has recently been demonstrated~\cite{Peterson2019PRL}. These progresses motivate people to pursue a more interesting but difficult task: observation of optomechanical effects at the single-photon level. This corresponds to a new parameter space in cavity optomechanical systems, namely the single-photon strong-coupling regime, in which there exist many interesting effects, such as photon blockade and photon correlation~\cite{Rabl2011PRL,Liao2013PRAc}, phonon sideband spectrum~\cite{Nunnenkamp2011PRL,Liao2012PRA}, and macroscopic quantum coherence~\cite{Marshall2003PRL,Liao2016PRL}.

Most previous studies in cavity optomechanics are focused on either the first-order or quadratic optomechanical systems. However, the mixed cavity optomechanical model~\cite{Rocheleau2010Nature,Xuereb2013PRA,Zhang2014PRA,Hauer2018PRA,Zhang2018PRA,Brunelli2018PRA,Zhou2019CTP,Sainadh2020OL} (\textit{with both the first-order and quadratic optomechanical couplings}) is a more basic model, which has attracted much recent attention from the community of optomechanics. There have been lots of works about the mixed cavity optomechanical model, including squeezing and cooling~\cite{Xuereb2013PRA}, self-sustained oscillations~\cite{Zhang2014PRA}, quantum nondemolition measurements of phonons~\cite{Hauer2018PRA}, optomechanically induced transparency~\cite{Zhang2018PRA}, preparation of nonclassical states~\cite{Brunelli2018PRA}, spectrum of single-photon emission and scattering~\cite{Zhou2019CTP}, and detection of weak forces~\cite{Sainadh2020OL}.

Though great effort has been devoted to the studies of both the single-photon strong optomechanical coupling and mixed optomechanical coupling, the single-photon strong-coupling or ultrastrong-coupling regime~\cite{Hu2015PRA,Garziano2015PRA,Macri2016PRA,Liao2020PRA,Sedov2020PRL} has not been experimentally realized in mixed optomechanical systems. As a result, how to implement an ultrastrong mixed-optomechanical interaction becomes a very interesting task. Quantum simulation~\cite{Buluta2009Science,Georgescu2014RMP,Eichler2018PRL}, as a state-of-the-art technique, might be a powerful way to explore the optomechanical interactions in the single-photon strong-coupling regime. Recently, a quantum simulation of mechano-optics has been proposed based on the quadratic optomechanical model~\cite{Bruschi2018NJP}. Note that some methods have been proposed to enhance the single-photon optomechanical couplings~\cite{Xuereb2012,Rimberg2014,Heikkila2014,Pirkkalainen2015,Johansson2014,Kim2015,Liao2014,Liao2015,Lue2015,Lemonde2016,Li2016,Yin2018,Liao2020PRA,Wang2017}. These methods include the coupling amplification effect via introducing the collective mechanical modes~\cite{Xuereb2012}, the sinusoidal modulation of optomechanical-coupling strength~\cite{Liao2014} or cavity-field frequency~\cite{Liao2015}, the utilizing of strong nonlinearity in the Josephson junctions~\cite{Rimberg2014,Heikkila2014,Pirkkalainen2015,Johansson2014,Kim2015,Bothner2020Naturephysics}, the coupling amplification via the squeezing~\cite{Lue2015,Lemonde2016,Li2016,Yin2018} or a displacement~\cite{Liao2020PRA} enhancement, and the utilizing of delayed quantum feedback control~\cite{Wang2017}.

In this work, we propose a scheme to realize this task by simulating a mixed optomechanical coupling in a coupled two-mode system, in which the two modes are coupled with each other via the cross-Kerr interaction. By introducing both single- and two-excitation drivings to one of the two modes, we can obtain the tunable and ultrastrong mixed-optomechanical interactions. In particular, the first-order (quadratic) optomechanical coupling strength could be greater than the effective frequency of the mechanical-like mode under proper conditions. Therefore, this effective mixed optomechanical model can enter the single-photon strong-coupling even ultrastrong-coupling regimes. As an application of the ultrastrong mixed-optomechanical couplings, we study how to generate the Schr\"{o}dinger cat states in the mechanical-like mode. We also investigate the quantum coherence effect in the generated states by calculating their Wigner functions.

The remaining part of this paper is organized as follows. In Sec.~\ref{sec2}, we show the physical model and the Hamiltonian. In Sec.~\ref{sec3}, we derive an approximate Hamiltonian for the mixed optomechanical interactions and evaluate its validity. In Sec.~\ref{sec4}, we study the preparation of the Schr\"{o}dinger cat states in mode $b$ and calculate their Wigner functions. In Sec.~\ref{sec5}, we present some discussions concerning the experimental implementation. Finally, we present a brief conclusion in Sec.~\ref{sec6}.

\section{Model and Hamiltonian\label{sec2}}

\begin{figure}[tbp]
\center
\includegraphics[bb=15 398 518 721, width=0.6 \textwidth]{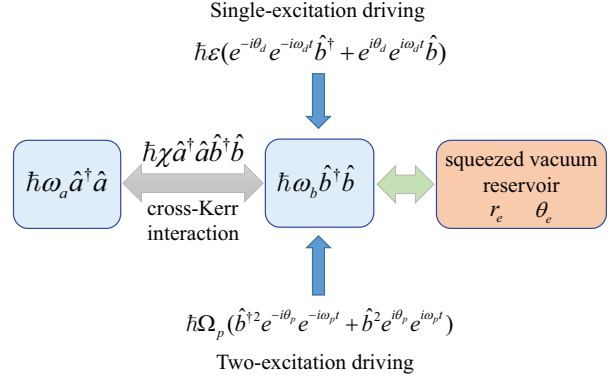}
\caption{Schematic diagram of the coupled two-bosonic-mode system. Two bosonic modes $a$ and $b$ (with corresponding resonance frequencies $\omega_{a}$ and $\omega_{b}$) are coupled by the cross-Kerr interaction (with the coupling strength $\chi$). Mode $b$ is subjected to both single-excitation driving (with phase $\theta_{d}$, frequency $\omega_{d}$, and amplitude $\varepsilon$) and two-excitation driving (with phase $\theta_{p}$, frequency $\omega_{p}$, and amplitude $\Omega_{p}$). A squeezed vacuum bath with the squeezing parameter $r_{e}$ and reference phase $\theta_{e}$ is introduced to mode $b$ for suppressing the excitations caused by the two-excitation driving.}
\label{mf}
\end{figure}

We consider a coupled two-bosonic-mode ($a$ and $b$) system (see Fig.~\ref{mf}), where the two modes are coupled with each other via a cross-Kerr interaction. In this system, one (e.g., mode $b$) of the two modes is subjected to both single- and two-excitation drivings. The Hamiltonian of the system reads
\begin{eqnarray}
\hat{H} &=&\hbar\omega_{a}\hat{a}^{\dagger}\hat{a}+\hbar\omega_{b}\hat{b}^{\dagger}\hat{b}+\hbar\chi\hat{a}^{\dagger}\hat{a}\hat{b}^{\dagger}%
\hat{b}+\hbar\varepsilon(e^{-i\theta_{d}}e^{-i\omega_{d}t}\hat{b}^{\dagger}+e^{i\theta_{d}}e^{i\omega_{d}t}\hat{b})  \notag \\
&&+\hbar\Omega_{p}(e^{-i\theta_{p}}e^{-i\omega_{p}t}\hat{b}^{\dagger2}+e^{i\theta_{p}}e^{i\omega_{p}t}\hat{b}^{2}),  \label{sdq1}
\end{eqnarray}
where $\hat{a}$ ($\hat{b}$) is the annihilation operator of the bosonic mode $a$ ($b$) with the resonance frequency $\omega_{a}$ ($\omega_{b}$). The third term in Eq.~(\ref{sdq1}) describes the cross-Kerr interaction with the coupling strength $\chi$. The parameters $\varepsilon e^{-i\theta_{d}}$ and $\omega_{d}$ ($\Omega_{p}e^{-i\theta_{p}}$ and $\omega_{p}$) are, respectively, the driving amplitude and frequency of the single (two)-excitation driving, with $\theta_{d}$ ($\theta_{p}$) being the driving phase.
In our model, the cross-Kerr interaction describes a general coupling between two bosonic modes~\cite{Hu2011PRA}. In principle, the involving two modes could be implemented by either two electromagnetic fields or two mechanical modes, and even one electromagnetic field and one mechanical mode. Note that some proposals have been proposed to implement the optomechanical couplings based on two microwave fields~\cite{Bothner2020Naturephysics,Johansson2014,Kim2015}. In this case, though the mechanical-like mode cannot be used to transduce other physical signals for sensing, the generated optomechanical interactions can be used to demonstrate the optomechanical physical effects. The two-excitation driving process can be realized with the degenerate parametric-down-conversion mechanism. In optical system, this process can be implemented by introducing the optical parametric amplifier~\cite{Walls2008book}. In circuit-QED systems, the two-excitation driving can be induced by a cycle three-level system~\cite{Wang2015PRA}, and this interaction has been widely used in various schemes~\cite{Zhao2020,Lue2015,Lemonde2016,Yin2018,Boissonneault2012PRA,Zhu2013PRB,Zhao2018PRA}.

In a rotating frame defined by $\exp(-i\hat{H}_{0}t/\hbar)$ with $\hat{H}_{0}=\hbar\omega_{a}\hat{a}^{\dagger}\hat{a}+\hbar\omega_{d}\hat{b}^{\dagger}\hat{b}$, the Hamiltonian $\hat{H}$ takes the form as
\begin{equation}
\hat{H}_{I}=\hbar\Delta_{b}\hat{b}^{\dagger}\hat{b}+\hbar\chi\hat{a}^{\dagger}\hat{a}\hat{b}^{\dagger}\hat{b}+\hbar\varepsilon(e^{-i\theta
_{d}}\hat{b}^{\dagger}+e^{i\theta_{d}}\hat{b})+\hbar\Omega_{p}(e^{-i\theta_{p}}\hat{b}^{\dagger 2}+e^{i\theta_{p}}\hat{b}^{2}),
\end{equation}
where $\Delta_{b}=\omega_{b}-\omega_{d}$ is the single-excitation driving detuning. Hereafter, we consider the case of $\omega_{p}=2\omega_{d}$.

To include the dissipations in this system, we assume that mode $a$ is connected with a heat bath and mode $b$ is in contact with a squeezed vacuum reservoir~\cite{You2018PRA} with the central frequency $\omega_{d}$. In this case, the evolution of the system in the rotating frame is governed by the quantum master equation
\begin{eqnarray}
\dot{\hat{\rho}} &=&-\frac{i}{\hbar}[\hat{H}_{I},\hat{\rho}]+\kappa_{a}(\bar{n}_{\text{th}}+1)\mathcal{D}[\hat{a}]\hat{\rho}+\kappa_{a}\bar{n}_{%
\text{th}}\mathcal{D}[\hat{a}^{\dagger}]\hat{\rho}+\kappa_{b}(N+1)\mathcal{D}[\hat{b}]\hat{\rho}+\kappa_{b}N\mathcal{D}[\hat{b}^{\dagger}]\hat{\rho}
\notag \\
&&-\kappa_{b}M\mathcal{G}[\hat{b}]\hat{\rho}-\kappa_{b}M^{\ast}\mathcal{G}[\hat{b}^{\dagger}]\hat{\rho}.  \label{mseq}
\end{eqnarray}
Here, $\mathcal{D}[\hat{o}]\hat{\rho} =\hat{o}\hat{\rho}\hat{o}^{\dagger}-(\hat{o}^{\dagger}\hat{o}\hat{\rho}+\hat{\rho}\hat{o}^{\dagger}\hat{o})/2$ and $\mathcal{G}[\hat{o}]\hat{\rho} =\hat{o}\hat{\rho}\hat{o}-(\hat{o}\hat{o}\hat{\rho}+\hat{\rho}\hat{o}\hat{o})/2$ are superoperators acting on the density matrix $\hat{\rho}$ of the system, $\kappa_{a}$ and $\kappa_{b}$ are the decay rates of modes $a$ and $b$, respectively, $\bar{n}_{\text{th}}$ is the thermal excitation number associated with mode $a$, $N=\sinh^{2}r_{e}$ is the mean photon number of the squeezed vacuum reservoir, and $M=\cosh r_{e}\sinh r_{e} e^{-i\theta_{e}}$ describes the strength of two-photon correlation~\cite{Scully1997book}, with $r_{e}$ being the squeezing parameter and $\theta_{e}$ the reference phase for the squeezed field.

\section{Quantum simulation for a mixed optomechanical interaction\label{sec3}}

In this section, we derive a mixed optomechanical Hamiltonian based on this driven two-mode system and evaluate the validity of the approximate Hamiltonian.

\subsection{Derivation of the mixed optomechanical Hamiltonian}

To exhibit the scheme for simulating the mixed optomechanical interaction between modes $a$ and $b$, we perform both the squeezing and displacement transformations to Eq.~(\ref{mseq}) by  $\hat{\rho}^{\prime}=\hat{D}_{b}^{\dagger}(\alpha)\hat{S}_{b}^{\dagger}(\zeta)\hat{\rho}\hat{S}_{b}(\zeta)\hat{D}_{b}(\alpha)$, where $\hat{D}_{b}(\alpha) =\exp(\alpha\hat{b}^{\dagger}-\alpha^{\ast}\hat{b})$ is the displacement operator and $\hat{S}_{b}(\zeta)=\exp[(\zeta^{\ast}\hat{b}^{2}-\zeta\hat{b}^{\dagger2})/2]$ is the squeezing operator with $\zeta =r\exp(i\varphi)$. In the transformed representation, the quantum master equation~(\ref{mseq}) becomes
\begin{eqnarray}
\dot{\hat{\rho}}^{\prime} &=&-\frac{i}{\hbar}[\hat{H}^{\prime},\hat{\rho}^{\prime}]+\kappa_{a}(\bar{n}_{\text{th}}+1)\mathcal{D}(\hat{a})\hat{\rho}^{\prime}+\kappa_{a}\bar{n}_{\text{th}}\mathcal{D}(\hat{a}^{\dagger})\hat{\rho}^{\prime}    \notag \\
&&+\kappa_{b}[\mathcal{N}(t)+1]\mathcal{D}(\hat{b})\hat{\rho}^{\prime}+\kappa_{b}\mathcal{N}(t)\mathcal{D}(\hat{b}^{\dagger})\hat{\rho}^{\prime}    \notag \\
&&-\kappa_{b}\mathcal{M}(t)\mathcal{G}(\hat{b})\hat{\rho}^{\prime}-\kappa_{b}\mathcal{M}^{\ast}(t)\mathcal{G}(\hat{b}^{\dagger})\hat{\rho}^{\prime}.  \label{timemasterequation}
\end{eqnarray}
Here, the effective thermal occupation $\mathcal{N}(t)$ and two-photon-correlation strength $\mathcal{M}(t)$ in the transformed representation are defined by
\begin{subequations}
\label{nmparameter}
\begin{align}
\mathcal{N}(t)=&\ \sinh^{2}(r_{e}-r)+\frac{1}{2}\cos^{2}\left[\frac{1}{2}(\varphi-\theta_{e})\right]\{\cosh[2(r_{e}+r)]-\cosh[2(r_{e}-r)]\},  \\
\mathcal{M}(t)=&\ \frac{1}{2}e^{-i\varphi}\sinh[2(r_{e}+r)]\cos^{2}\left[\frac{1}{2}(\varphi-\theta_{e})\right] -\frac{1}{2}e^{-i\varphi
}\sinh [2(r_{e}-r)]\sin^{2}\left[\frac{1}{2}(\varphi-\theta_{e})\right]      \notag \\
&+\frac{i}{2}e^{-i\varphi}\sin(\varphi-\theta_{e})\sinh(2r_{e}).
\end{align}
\end{subequations}
The transformed Hamiltonian, apart from a ``c"-number term, reads
\begin{eqnarray}
\hat{H}^{\prime} &=&\hbar\tilde{\omega}_{a}\hat{a}^{\dagger}\hat{a}+\hbar\tilde{\omega}_{b}\hat{b}^{\dagger}\hat{b}+\hbar\text{Re}(\alpha
e^{-i\varphi/2})\chi e^{-2r}\hat{a}^{\dagger}\hat{a}(\hat{b}^{\dagger}e^{i\varphi/2}+\hat{b}e^{-i\varphi/2})  \notag \\
&&+i\hbar\text{Im}(\alpha e^{-i\varphi/2})\chi e^{2r}\hat{a}^{\dagger}\hat{a}(\hat{b}^{\dagger}e^{i\varphi/2}-\hat{b}e^{-i\varphi/2})+\hbar
\frac{\chi}{4}e^{-2r}\hat{a}^{\dagger}\hat{a}(\hat{b}^{\dagger}e^{i\varphi/2}+\hat{b}e^{-i\varphi/2})^{2}  \notag \\
&&-\hbar\frac{\chi}{4}e^{2r}\hat{a}^{\dagger}\hat{a}(\hat{b}^{\dagger}e^{i\varphi/2}-\hat{b}e^{-i\varphi/2})^{2},   \label{timehamil}
\end{eqnarray}
where $\text{Re}(\alpha e^{-i\varphi /2})$ and $\text{Im}(\alpha e^{-i\varphi /2})$ denote, respectively, the real and imagine parts of the argument $\alpha e^{-i\varphi /2}$, and the effective frequencies of modes $a$ and $b$ are defined by
\begin{subequations}
\label{effq}
\begin{align}
\tilde{\omega}_{a}=&\ \chi\{e^{-2r}[\text{Re}(\alpha e^{-i\varphi/2})]^{2}+e^{2r}[\text{Im}(\alpha e^{-i\varphi /2})]^{2}\}-\frac{\chi}{2}, \\
\tilde{\omega}_{b}=&\ \Delta_{b}\cosh(2r)-2\Omega_{p}\cos(\theta_{p}+\varphi)\sinh(2r)+\dot{\varphi}\sinh^{2}r.
\end{align}
\end{subequations}
The parameters $r$, $\varphi$, and $\alpha$ in Eqs.~(\ref{nmparameter}-\ref{effq}) are determined by the equations
\begin{subequations}
\label{paequation}
\begin{align}
\dot{r}=&\ 2\Omega_{p}\sin(\theta_{p}+\varphi),   \label{requ} \\
\dot{\varphi}=&\ 4\Omega_{p}\coth(2r)\cos(\theta_{p}+\varphi)-2\Delta_{b},   \label{phequ} \\
\dot{\alpha}=& -\frac{\kappa_{b}}{2}\alpha-i[\Delta_{b}\cosh(2r)-2\Omega_{p}\cos(\theta_{p}+\varphi)\sinh(2r)  \notag \\
& +\dot{\varphi}\sinh^{2}r]\alpha-i\varepsilon\lbrack e^{-i\theta_{d}}\cosh r-e^{i(\theta_{d}+\varphi)}\sinh r].
\end{align}
\end{subequations}

It can be seen from Eq.~(\ref{paequation}) that the equations of $r$ and $\varphi$ are independent of the decay rates of the system. Below we analyze Eqs.~(\ref{requ}) and (\ref{phequ}) firstly. For a chosen two-excitation driving, the parameters $\Omega_{p}$ and $\theta_{p}$ are given. Then the value of $r$ and $\varphi$ can be determined by Eqs.~(\ref{requ}) and (\ref{phequ}). To obtain a stationary quadratic optomechanical coupling, we consider a case of $\theta_{p}=\pi$ and $\varphi=\pi$, which leads to
\begin{equation}
r=\frac{1}{4}\ln\left(\frac{\Delta_{b}+2\Omega_{p}}{\Delta_{b}-2\Omega_{p}}\right).     \label{squeezingparameter}
\end{equation}
In addition, we only study the steady-state displacement case where the time scale of system relaxation is much shorter than other time scales. Then the steady-state displacement amplitude is obtained as
\begin{equation}
\alpha_{\text{ss}}=\frac{\varepsilon e^{-r}}{\sqrt{(\Delta_{b}-2\Omega_{p})^{2}+\kappa_{b}^{2}/4}}. \label{daphequation}
\end{equation}
Hereafter, we consider the case of $\tan\theta_{d}=\kappa_{b}/[2(\Delta_{b}-2\Omega_{p})]$ so that $\alpha_{\text{ss}}$ is a real number for simplicity.

\begin{figure}[tbp]
\center
\includegraphics[bb=-2 0 503 480, width=0.68 \textwidth]{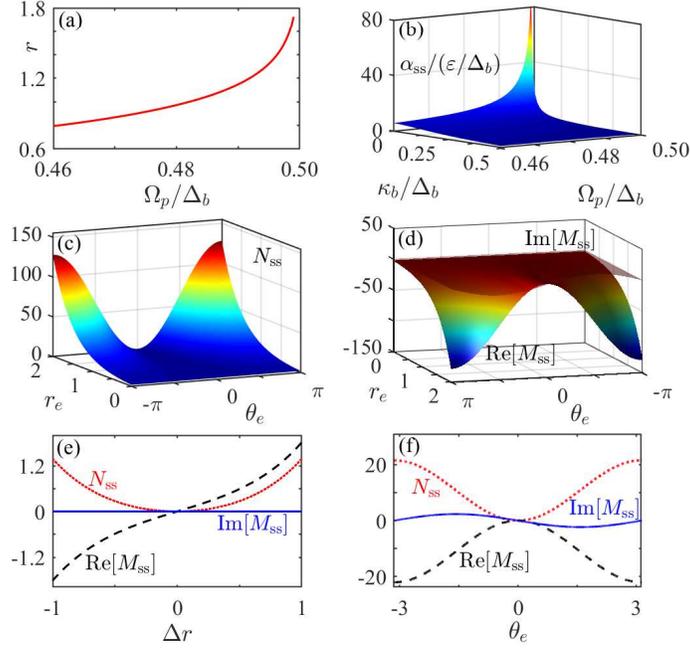}
\caption{(a) The squeezing amplitude $r$ as a function of $\Omega_{p}/\Delta_{b}$. (b) The scaled displacement amplitude $\alpha_{\text{ss}}/(\varepsilon/\Delta_{b})$ as a function of $\kappa_{b}/\Delta_{b}$ and $\Omega_{p}/\Delta_{b}$. (c) $N_{\text{ss}}$ as a function of $r_{e}$ and $\theta_{e}$ with the parameter $r=1.12$. (d) Re$[M_{\text{ss}}]$ and Im$[M_{\text{ss}}]$ as functions of $r_{e}$ and $\theta_{e}$ with the parameter $r=1.12$. (e) $N_{\text{ss}}$ (dotted red curve), Re$[M_{\text{ss}}]$ (dashed black curve), and Im$[M_{\text{ss}}]$ (solid blue curve) versus $\Delta r=r_{e}-r$ at $\theta_{e}=0$. (f) $N_{\text{ss}}$ (dotted red curve), Re$[M_{\text{ss}}]$ (dashed black curve), and Im$[M_{\text{ss}}]$ (solid blue curve) versus $\theta_{e}$ at $r_{e}=r=1.12$.}
\label{effectivethermalnoise}
\end{figure}

In Fig.~\ref{effectivethermalnoise}(a), we show the squeezing amplitude $r$ defined in Eq.~(\ref{squeezingparameter}) as a function of $\Omega_{p}/\Delta_{b}$. Here, we can find that $r$ increases with the increase of the ratio $\Omega_{p}/\Delta_{b}$. In particular, the value of $r$ could be increased by choosing $\Delta_{b}-2\Omega_{p}\simeq0$ such that both the first-order and quadratic optomechanical interactions can be enhanced largely. This point is very important for the realization of single-photon strong-coupling regime for the optomechanical interactions.
In Fig.~\ref{effectivethermalnoise}(b), we show the scaled displacement amplitude $\alpha_{\text{ss}}/(\varepsilon/\Delta_{b})$ as a function of $\kappa_{b}/\Delta_{b}$ and $\Omega_{p}/\Delta_{b}$ based on Eq.~(\ref{daphequation}). Here, we can see that the displacement amplitude $\alpha_{\text{ss}}$ is proportional to the single-excitation driving amplitude $\varepsilon$. For a given $\varepsilon$, we can choose a small $\kappa_{b}$ and a proper value of $\Omega_{p}/\Delta_{b}$ (approaching $1/2$) to get a large $\alpha_{\text{ss}}$. In addition, the point $\Delta_{b}=2\Omega_{p}$ is not a singularity in the presence of dissipation $\kappa_{b}$.

Based on the above analyses, we know that the system, under the steady-state displacement $\alpha_{\text{ss}}$, is governed by the following quantum master equation,
\begin{eqnarray}
\dot{\hat{\rho}}^{\prime} &=&-\frac{i}{\hbar}[\hat{H}^{\prime},\hat{\rho}^{\prime}]+\kappa_{a}(\bar{n}_{\text{th}}+1)\mathcal{D}[\hat{a}]\hat{\rho}^{\prime}+\kappa_{a}\bar{n}_{\text{th}}\mathcal{D}[\hat{a}^{\dagger}]\hat{\rho}^{\prime}  \notag \\
&&+\kappa_{b}(N_{\text{ss}}+1)\mathcal{D}[\hat{b}]\hat{\rho}^{\prime}+\kappa_{b}N_{\text{ss}}\mathcal{D}[\hat{b}^{\dagger}]\hat{\rho}^{\prime}  \notag \\
&&-\kappa_{b}M_{\text{ss}}\mathcal{G}[\hat{b}]\hat{\rho}^{\prime}-\kappa_{b}M_{\text{ss}}^{\ast}\mathcal{G}[\hat{b}^{\dagger}]\hat{\rho}^{\prime},
\label{masterequation}
\end{eqnarray}
where the transformed Hamiltonian $\hat{H}^{\prime}$ becomes
\begin{equation}
\hat{H}^{\prime}=\hbar\tilde{\omega}_{a}\hat{a}^{\dagger}\hat{a}+\hbar\tilde{\omega}_{b}\hat{b}^{\dagger}\hat{b}+\hbar g_{1}\hat{a}^{\dagger}%
\hat{a}(\hat{b}^{\dagger}+\hat{b})+\hbar g_{2}\hat{a}^{\dagger}\hat{a}(\hat{b}^{\dagger}+\hat{b})^{2}-\hbar g_{2}^{\prime}\hat{a}^{\dagger}\hat{a%
}(\hat{b}^{\dagger}-\hat{b})^{2},  \label{Hamiltonianprime}
\end{equation}
and the effective frequencies of modes $a$ and $b$ are reduced to $\tilde{\omega}_{a}=-\chi/2+\chi\alpha_{\text{ss}}^{2}e^{2r}$ and $\tilde{\omega}_{b}=(\Delta_{b}-2\Omega_{p}) e^{2r}=\Delta_{b}-2\Omega_{p}\sinh r/\cosh r$.
Those effective coupling strengthes in Hamiltonian $\hat{H}^{\prime}$ are given by
\begin{equation}
g_{1}=\chi e^{2r}\alpha_{\text{ss}},\hspace{0.5cm}g_{2}=\frac{\chi}{4}e^{2r},\hspace{0.5cm}g_{2}^{\prime}=\frac{\chi}{4}e^{-2r}.  \label{mixcoupling}
\end{equation}
We can see from Eqs.~(\ref{squeezingparameter}) and (\ref{daphequation}) that the coupling strengths of both the first-order and quadratic optomechanical interactions can be tuned by controlling the single- and two-excitation driving parameters. Meanwhile, the value of $\alpha_{\text{ss}}$ could be either larger or smaller than $1/4$ which means that the first-order optomechanical coupling could be either stronger or weaker than the quadratic optomechanical coupling.

In Eq.~(\ref{masterequation}), the parameters $N_{\text{ss}}$ and $M_{\text{ss}}$ are, respectively, the effective thermal occupation number and two-photon-correlation strength in the transformed representation, which take the form
\begin{subequations}
\label{paens}
\begin{align}
N_{\text{ss}}=&\ \frac{1}{2}\{\cosh [2(r_{e}+r)]-\cosh [2(r_{e}-r)]\}\sin^{2}\left( \frac{\theta _{e}}{2}\right) +\sinh ^{2}(r_{e}-r), \\
M_{\text{ss}}=&\ \frac{1}{2}\left\{\sinh [2(r_{e}-r)]\cos^{2}\left(\frac{\theta_{e}}{2}\right)-\sinh[2(r_{e}+r)]\sin^{2}\left(\frac{\theta_{e}}{%
2}\right)\right\}-\frac{i}{2}\sin(\theta_{e})\sinh(2r_{e}).
\end{align}
\end{subequations}
It can be seen from Eq.~(\ref{paens}) that the parameters $N_{\text{ss}}$ and $M_{\text{ss}}$ depend on the three parameters $\theta_{e}$, $r_{e}$, and $r$. In principle, when we take a given $r$, then we can plot the parameters $N_{\text{ss}}$ and $M_{\text{ss}}$ as functions of $\theta_{e}$ and $r_{e}$, as shown in Figs.~\ref{effectivethermalnoise}(c) and \ref{effectivethermalnoise}(d). Here, we can see that $N_{\text{ss}}$ and $M_{\text{ss}}$ become very large in a large parameter range. In the vicinity of $\theta_{e}=0$ and $r_{e}=r$, however, the values of $N_{\text{ss}}$ and $M_{\text{ss}}$ are very small. In order to understand the nature of $N_{\text{ss}}$ and $M_{\text{ss}}$ more clearly, we consider two interesting cases. (i) When $\theta_{e}=0$, we plot $N_{\text{ss}}$ and $M_{\text{ss}}$ as a function of $\Delta r=r_{e}-r$ in Fig.~\ref{effectivethermalnoise}(e). We have $N_{\text{ss}}=0$ and $M_{\text{ss}}=0$ at the point $r_{e}=r$. (ii) When $r_{e}=r$, we plot $N_{\text{ss}}$ and $M_{\text{ss}}$ as a function of $\theta_{e}$ in Fig.~\ref{effectivethermalnoise}(f). We also see $N_{\text{ss}}=0$ and $M_{\text{ss}}=0$ at the point $\theta_{e}=0$. It should be emphasized that $N_{\text{ss}}=0$ and $M_{\text{ss}}=0$ imply that both the thermal noise and the squeezing vacuum noise have been suppressed completely~\cite{Lue2015}. This working point is very important to observe the optomechanical effects at the single-photon level. This feature is also the motivation for introducing the squeezing bath, i.e., using a well-designed squeezing vacuum bath to suppress the thermal noise and the excitations caused by the two-excitation driving.
\begin{figure}[tbp]
\center
\includegraphics[bb=15 14 711 433, width=0.6 \textwidth]{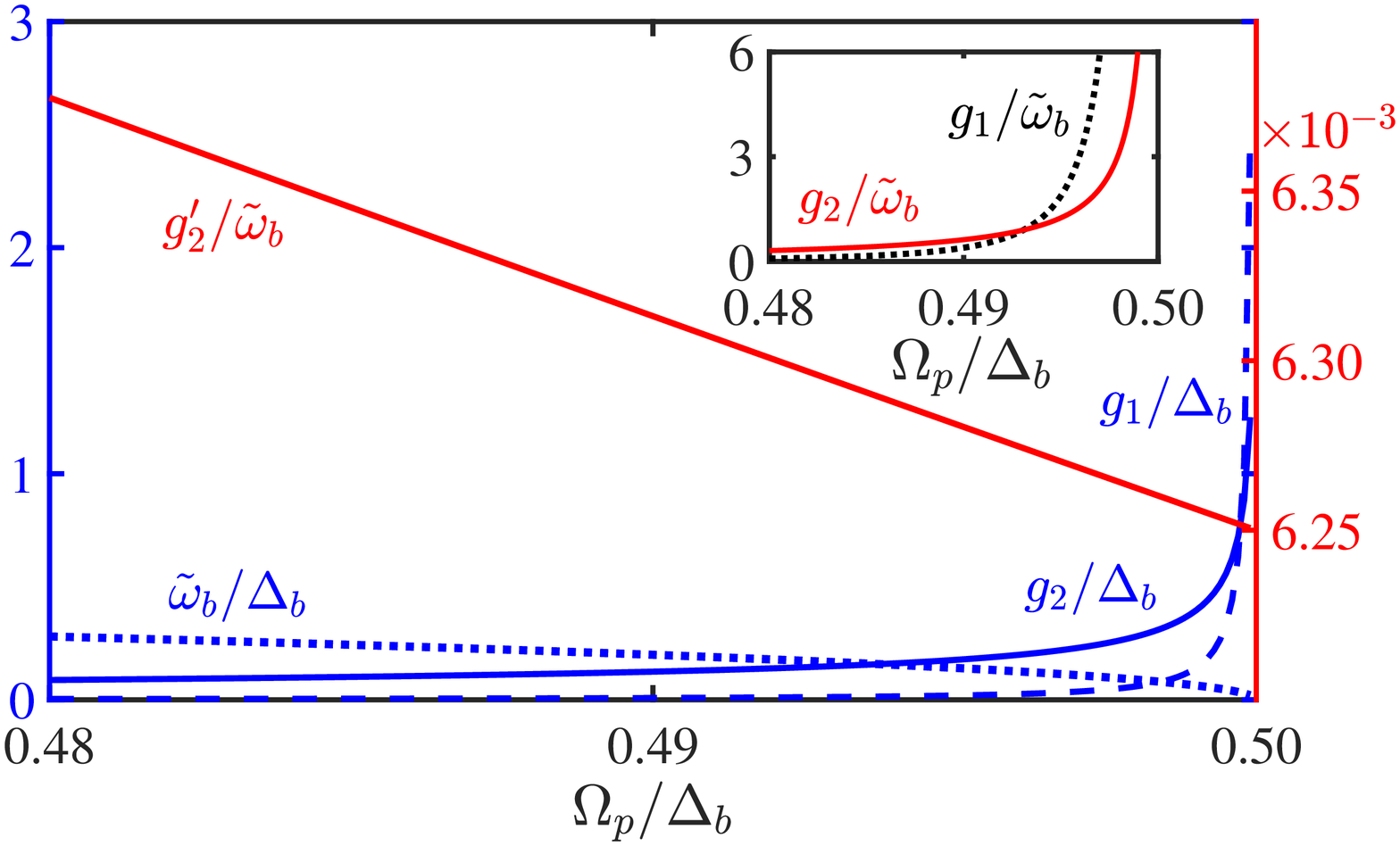}
\caption{Plots of $\tilde{\protect\omega}_{b}/\Delta_{b}$ (dotted blue line), $g_{1}/\Delta_{b}$ (dashed blue line), $g_{2}/\Delta_{b}$ (solid blue line), and $g_{2}^{\prime}/\tilde{\omega}_{b}$ (solid red line) as functions of $\Omega_{p}/\Delta_{b}$. In the inset, we plot $g_{1}/\tilde{\protect\omega}_{b}$ and $g_{2}/\tilde{\protect\omega}_{b}$ as functions of $\Omega_{p}/\Delta_{b}$. Other parameters used are $\chi/\Delta_{b}=0.05$, $\varepsilon/\Delta_{b}=0.001$, and $\theta_{d}=0$.}
\label{parameters}
\end{figure}

In this paper, we focus on the red-detuning driving case, i.e., $\Delta_{b}>0$. It can be seen from Eq.~(\ref{squeezingparameter}) that the squeezing amplitude $r$ will become very large at $2\Omega_{p}\approx\Delta_{b}$. Under the parameter conditions $g_{2}\gg g_{2}^{\prime}$ and $\tilde{\omega}_{b}\gg 2n_{\text{max}}g_{2}^{\prime}$ with $n_{\text{max}}$ being the largest photon number in mode $a$, the term $-\hbar g_{2}^{\prime}\hat{a}^{\dagger}\hat{a}(\hat{b}^{\dagger2}+\hat{b}^{2})$ can be safely neglected under the rotating-wave approximation. Meanwhile, we can safely discard the cross-Kerr term $2\hbar g_{2}^{\prime}\hat{a}^{\dagger}\hat{a}\hat{b}^{\dagger}\hat{b}$ because the frequency shift caused by this term for mode $b$ is much smaller than the effective frequency of mode $b$. Then Eq.~(\ref{Hamiltonianprime}) is reduced to
\begin{equation}
\hat{H}_{\text{app}}=\hbar\tilde{\omega}_{a}^{\prime}\hat{a}^{\dagger}\hat{a}+\hbar\tilde{\omega}_{b}\hat{b}^{\dagger}\hat{b}+\hbar g_{1}\hat{a}%
^{\dagger}\hat{a}(\hat{b}^{\dagger}+\hat{b})+\hbar g_{2}\hat{a}^{\dagger}\hat{a}(\hat{b}^{\dagger}+\hat{b})^{2},   \label{mixedcoptomechanicalHamiltonian}
\end{equation}
where $\tilde{\omega}_{a}^{\prime}=\tilde{\omega}_{a}+g_{2}^{\prime}$ denotes the renormalized frequency of mode $a$. The approximate Hamiltonian $\hat{H}_{\text{app}}$ describes the standard mixed optomechanical model consisting of both the first-order and quadratic optomechanical interactions, where modes $a$ and $b$ play the role of optical mode and mechanical mode, with the effective resonance frequencies $\tilde{\omega}_{a}^{\prime}$ and $\tilde{\omega}_{b}$, respectively. The parameters $g_{1}$ and $g_{2}$ are, respectively, the single-photon first-order and quadratic optomechanical coupling strengths.
\begin{figure}[tbp]
\center
\includegraphics[bb=0 0 435 358, width=0.72 \textwidth]{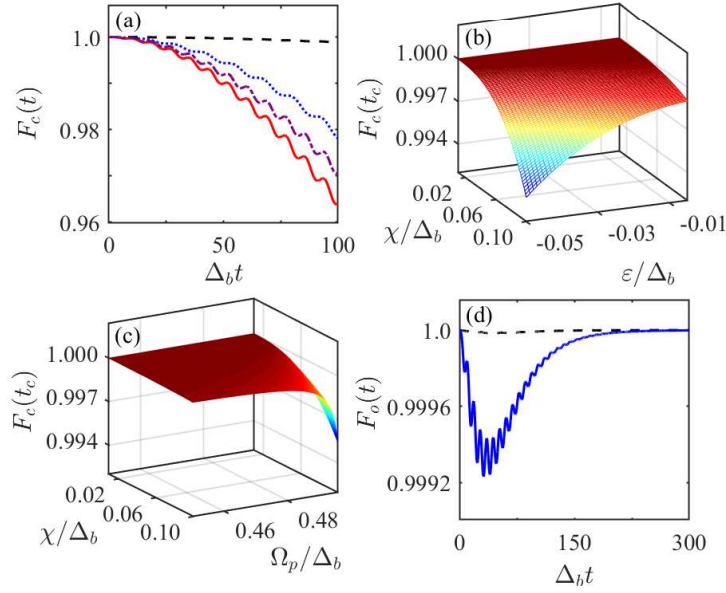}
\caption{(a) Dynamics of the fidelity $F_{c}(t)$ in the closed-system case when the parameters $\chi/\Delta_{b}$ and $\varepsilon/\Delta_{b}$ take different values: $\chi/\Delta_{b}=0.01$ and $\varepsilon/\Delta_{b}=0.1$ (dashed black curve), $\chi/\Delta_{b}=0.01$ and $\varepsilon/\Delta_{b}=0.5$ (dotted blue curve), $\chi/\Delta_{b}=0.05$ and $\varepsilon/\Delta_{b}=0.005$ (dot-dashed purple), and $\chi/\Delta_{b}=0.05$ and $\varepsilon/\Delta_{b}=0.01$ (solid red curve). (b) The fidelity $F_{c}(t_{c})$ as a function of $\chi/\Delta_{b}$ and $\varepsilon/\Delta_{b}$ at $\Omega_{p}/\Delta_{b}=0.49$. (c) The fidelity $F_{c}(t_{c})$ as a function of $\chi/\Delta_{b}$ and $\Omega_{p}/\Delta_{b}$ at $\varepsilon/\Delta_{b}=0.01$. (d) The fidelity $F_{o}(t)$ in the open-system case as a function of $\Delta_{b}t$ under the parameters $\chi/\Delta_{b}=0.01$ and Re$[\varepsilon\exp(i\theta_{d})]/\Delta_{b}=0.05$ (dashed black curve), $\chi/\Delta_{b}=0.05$ and Re$[\varepsilon\exp(i\theta_{d})]/\Delta_{b}=0.01$ (solid blue curve), other parameters used are $\Omega_{p}/\Delta_{b}=0.49$, $r_{e}=r$, $\theta_{e}=0$, $\bar{n}_{\text{th}}=0$, $\kappa_{a}/\Delta_{b}=0.02$, and $\kappa_{b}/\Delta_{b}=0.02$. The initial state of the system is $|1\rangle_{a}|\beta\rangle_{b}$ with coherent state $|\beta\rangle$ ($\beta =0.1$) in panels (a) and (d), while the initial state of the system is $[(|0\rangle_{a}+|1\rangle_{a}) |0\rangle_{b}]/\sqrt{2}$ in panels (b) and (c).}
\label{Fidelity}
\end{figure}

To characterize these parameters in $\hat{H}_{\text{app}}$, we investigate the magnitude of these related parameters as functions of the detuning $\Delta_{b}$ of mode $b$. Concretely, Fig.~\ref{parameters} shows the parameter $\tilde{\omega}_{b}/\Delta_{b}$ as a function of $\Omega_{p}/\Delta_{b}$ (dotted blue line), we find that the effective frequency $\tilde{\omega}_{b}$ of mode $b$ decreases with the increase of driving strength $\Omega_{p}$. We also plot the ratios $g_{1}/\Delta_{b}$, $g_{2}/\Delta_{b}$, and $g_{2}^{\prime}/\tilde{\omega}_{b}$ as functions of $\Omega_{p}/\Delta_{b}$. Here, we can see that the coupling strengths $g_{1}$ and $g_{2}$ increase with the increase of the driving amplitude $\Omega_{p}$. In addition, we can see that the coupling strengths $g_{1}$ and $g_{2}$ could be a considerable fraction of (and smaller than) $\tilde{\omega}_{b}$ (see the inset), which means that both the first-order and quadratic optomechanical couplings could enter the ultrastrong-coupling regime~\cite{Kockum2019NatRevPhys}. In particular, when the coupling strengths $g_{1}$ and $g_{2}$ are larger than the resonance frequency $\tilde{\omega}_{b}$, the system enters the so-called  deep-strong coupling regime. As show in Fig.~\ref{parameters}, the ratio $g_{2}^{\prime}/\tilde{\omega}_{b}$ decreases with the increase of $\Omega_{p}/\Delta_{b}$, and its value is much smaller than $1$, which means that the two terms $2\hbar g_{2}^{\prime}\hat{a}^{\dagger}\hat{a}\hat{b}^{\dagger}\hat{b}$ and $-\hbar g_{2}^{\prime}\hat{a}^{\dagger}\hat{a}(\hat{b}^{\dagger2}+\hat{b}^{2})$ can be safely discarded.

Interestingly, when $\varepsilon=0$, $e^{4r}\gg 1$, and $\tilde{\omega}_{b}\gg 2n_{\text{max}}g_{2}^{\prime}$, we can obtain an approximate quadratic optomechanical Hamiltonian
\begin{equation}
\hat{H}_{\text{app}}^{\prime }=\hbar \tilde{\omega}_{a}^{\prime }%
\hat{a}^{\dagger }\hat{a}+\hbar \tilde{\omega}_{b}\hat{b}^{\dagger
}\hat{b}+\hbar g_{2}\hat{a}^{\dagger }\hat{a}(\hat{b}^{\dagger }+%
\hat{b})^{2}.  \label{quadraticoptomechanicalHamiltonian}
\end{equation}
Here, $g_{2}$ could reach a considerable fraction of $\tilde{\omega}_{b}$ under proper parameters, which means that the physical system corresponding to Eq.~(\ref{quadraticoptomechanicalHamiltonian}) can work in the single-photon strong-coupling even ultrastrong-coupling regimes of the quadratic optomechanical coupling.

\subsection{Evaluation of the validity of the approximate Hamiltonian $\hat{H}_{\text{app}}$}

To study the validity of the approximate Hamiltonian $\hat{H}_{\text{app}}$ given in Eq.~(\ref{mixedcoptomechanicalHamiltonian}), we adopt the fidelity between the approximate state $|\psi _{\text{app}}(t)\rangle$ and exact state $|\psi _{\text{ext}}(t)\rangle$, which are obtained with the approximate Hamiltonian (\ref{mixedcoptomechanicalHamiltonian}) and the exact Hamiltonian (\ref{Hamiltonianprime}), respectively. For avoiding the crosstalk from the system dissipations, we first study the closed-system case. In this case, the fidelity can be calculated by $F_{c}(t)=|\langle\psi_{\text{ext}}(t)|\psi_{\text{app}}(t)\rangle|$. In order to know the validity of the approximate Hamiltonians in the low-excitation regime, we choose the initial state as $|1\rangle_{a}|\beta\rangle_{b}$ with coherent state $|\beta\rangle$. In Fig.~\ref{Fidelity}(a), we plot the fidelity $F_{c}(t)$ as a function of the time $\Delta_{b}t$. Here, we can see that the fidelity $F_{c}(t)$ is high ($>0.96$) under the used parameters.
Considering that the system can produce the cat states at time $t_{c}=\pi/\sqrt{(\tilde{\omega}_{b}+4g_{2})\tilde{\omega}_{b}}$, we show the fidelity $F_{c}(t_{c})$ as a function of the parameters $\chi/\Delta_{b}$ and $\varepsilon/\Delta_{b}$ in Fig.~\ref{Fidelity}(b), and we also show the fidelity $F_{c}(t_{c})$ as a function of the parameters $\chi/\Delta_{b}$ and $\Omega_{p}/\Delta_{b}$ in Fig.~\ref{Fidelity}(c). We can see that the fidelity is high in a wide parameter space.

We also study the fidelity in the realistic case by including the dissipations. In the open-system case, the state of the system is described by the density matrix. The fidelity between the exact density matrix $\hat{\rho}_{\text{ext}}$ and the approximate density matrix $\hat{\rho}_{\text{app}}$ can be calculated by $F_{o}(t)=$Tr$[(\sqrt{\hat{\rho}_{\text{ext}}}\hat{\rho}_{\text{app}}\sqrt{\hat{\rho}_{\text{ext}}})^{1/2}]$, where the exact density matrix evolves under Eq.~(\ref{masterequation}) with the exact Hamiltonian, while the approximate density matrix is governed by Eq.~(\ref{masterequation}) under the replacement $\hat{H}^{\prime}\rightarrow\hat{H}_{\text{app}}$. In Fig.~\ref{Fidelity}(d), we show the dynamical evolution of the fidelity corresponding to Eq.~(\ref{mixedcoptomechanicalHamiltonian}). Here, we can see that, due to the dissipation, the fidelity of the open system approaches gradually to a stationary value in the long-time limit. Therefore, the approximate Hamiltonians $\hat{H}_{\text{app}}$ given by (\ref{mixedcoptomechanicalHamiltonian}) is valid in both the closed- and open-system cases.

\section{Generation of macroscopic quantum superposed states\label{sec4}}

One of the important applications of the simulated optomechanical interactions is the generation of macroscopic quantum superposed states. Now, we study the generation of a superposition of coherent squeezed state and vacuum state based on the mixed optomechanical interaction. To this end, we consider the initial state as $|\Psi(0)\rangle =[(|0\rangle_{a}+|1\rangle_{a})|0\rangle_{b}]/\sqrt{2}$, where $|0\rangle_{a}$ and $|1\rangle_{a}$ are, respectively, the vacuum state and single-photon state of mode $a$, while $|0\rangle_{b}$ is the vacuum state of mode $b$. To see the analytical expression of the generated states, in this section we analytically derive the state evolution in the closed-system case. We also calculate the Wigner function of the generated states in the presence of dissipations by numerically solving the quantum master equations.
Based on Eq.~(\ref{mixedcoptomechanicalHamiltonian}), the evolution of the system can be calculated, and the state of the system at time $t$ becomes
\begin{equation}
|\Psi (t)\rangle =\frac{1}{\sqrt{2}}\{|0\rangle_{a}|0\rangle_{b}+e^{-i\epsilon(t)}|1\rangle_{a}\hat{D}_{b}[\alpha_{1}(t)]\hat{S}%
_{b}(\eta_{1})\hat{S}_{b}(-\eta_{1}e^{-2i\varpi_{1}t})|0\rangle_{b}\},   \label{timestate}
\end{equation}
where we introduce the squeezing parameter $\eta_{1}=\ln [(\tilde{\omega}_{b}+4g_{2})/\tilde{\omega}_{b}]/4$ and the displacement parameter $\alpha_{1}(t)=\beta_{1}e^{-\eta_{1}}[1-\cos (\varpi_{1}t)+ie^{2\eta_{1}}\sin (\varpi_{1}t)]$ with $\beta_{1}=-g_{1}e^{-3\eta_{1}}/\tilde{\omega}_{b}$ and $\varpi_{1}=e^{2\eta_{1}}\tilde{\omega}_{b}$. The phase $\epsilon(t)$ in Eq.~(\ref{timestate}) is defined by $\epsilon(t)=\varepsilon_{1,0}t+\beta_{1}^{2}\sin(\varpi_{1}t)$
with $\varepsilon_{1,0}=\tilde{\omega}_{a}^{\prime}-g_{1}^{2}e^{-4\eta_{1}}/\tilde{\omega}_{b}+(\varpi_{1}-\tilde{\omega}_{b})/2$. According to Eq.~(\ref{timestate}), we know that the coherent-state component for mode $b$ can be generated at time $t_{c}=(2k+1)\pi/\varpi_{1}$ with natural numbers $k$. By expressing mode $a$ with the basis states $|\varphi_{\pm}\rangle_{a}=\left(|0\rangle_{a}\pm|1\rangle_{a}\right)/\sqrt{2}$, then the state of the system becomes
\begin{equation}
|\Psi(t_{c})\rangle =\frac{1}{2}\left[\frac{1}{N_{+}}|\varphi_{+}\rangle_{a}|\phi_{+}(t_{c})\rangle_{b}+\frac{1}{N_{-}}|\varphi_{-}\rangle
_{a}|\phi_{-}(t_{c})\rangle_{b}\right],
\end{equation}
where we introduced the cat states of mode $b$ as
\begin{equation}
|\phi_{\pm}(t_{c})\rangle_{b}=N_{\pm}[|0\rangle_{b}\pm e^{-i\varepsilon_{1,0}t_{c}}|\alpha_{1}(t_{c})\rangle_{b}],
\end{equation}
\begin{figure}[tbp]
\center
\includegraphics[width=0.98 \textwidth]{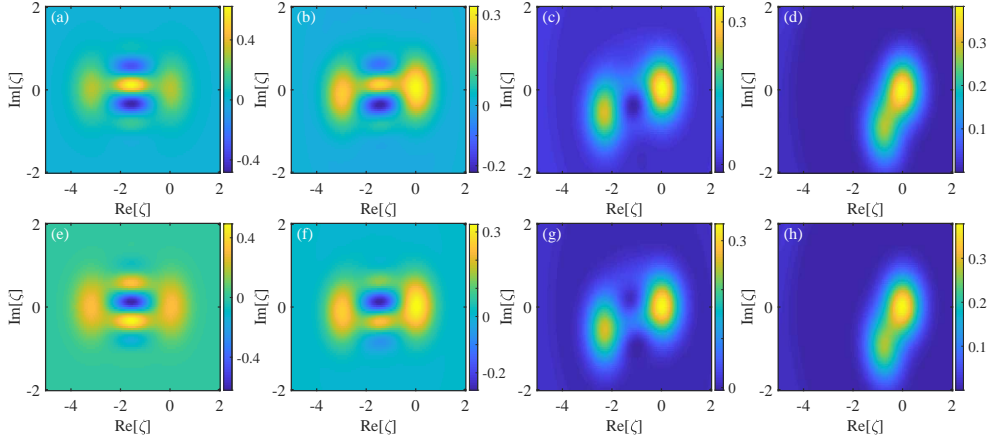}
\caption{The Wigner functions $W_{+}(\zeta)$ ($W_{-}(\zeta)$) of the generated states $\hat{\rho}_{b}^{(+)}(t_{c})$ ($\hat{\rho}_{b}^{(-)}(t_{c})$) are shown in the first (second) row at various decay rates: (a), (e) $\kappa_{a}/\Delta_{b}=\kappa_{b}/\Delta_{b}=0$; (b), (f) $\kappa_{a}/\Delta_{b}=\kappa_{b}/\Delta_{b}=0.02$; (c), (g) $\kappa_{a}/\Delta_{b}=0.02$ and $\kappa_{b}/\Delta_{b}=0.1$; and (d), (h) $\kappa_{a}/\Delta_{b}=0.02$ and $\kappa_{b}/\Delta_{b}=0.5$. Other parameters used are $t_{c}=\pi/\varpi_{1}$, $r_{e}=r$, $\theta_{e}=0$, $\bar{n}_{\text{th}}=0$, $\Omega_{p}/\Delta_{b}=0.49$, $\chi/\Delta_{b}=0.01$, and Re$[\varepsilon\exp(i\theta_{d})]/\Delta_{b}=0.3$ corresponding to $g_{1}/\tilde{\omega}_{b}=2.378$ and $g_{2}/\tilde{\omega}_{b}=0.125$.}
\label{wignercat}
\end{figure}
with the coherent state amplitude $\alpha_{1}(t_{c})=-2g_{1}/(4g_{2}+\tilde{\omega}_{b})$ and the normalization constants $N_{\pm}=\{{2\pm 2\exp [-\alpha _{1}^{2}(t_{c})/2]\cos(\varepsilon_{1,0}t_{c})\}}^{-1/2}$. Mode $b$ will collapse into the cat states $|\phi_{\pm}(t_{c})\rangle_{b}$ when the states $|\varphi_{\pm}\rangle_{a}$ are measured with the measuring probabilities $P_{\pm}(t_{c})=1/(4N_{+}^{2})$, respectively. For a relatively large $\alpha_{1}(t_{c})$, we have $P_{\pm}(t_{c})\approx1/2$.

The quantum coherence and interference in the generated states can be investigated by calculating the Wigner function~\cite{Barnett1997book}, which is defined by
\begin{equation}
W(\zeta) =\frac{2}{\pi}\text{Tr}[\hat{D}^{\dagger}(\zeta)\hat{\rho}\hat{D}(\zeta)(-1)^{\hat{b}^{\dagger}\hat{b}}],
\end{equation}
for the density matrix $\hat{\rho}$, where $\zeta$ is a complex variable. For the states $|\phi_{\pm}(t_{c})\rangle_{b}$, we show the Wigner functions $W_{\pm}(\zeta)$ in Figs.~\ref{wignercat}(a) and \ref{wignercat}(e), which exhibit clear quantum interference pattern and evidence of macroscopically distinct superposition components. To study the cat-state generation in the presence of system dissipations, in Figs.~\ref{wignercat}(b)-\ref{wignercat}(d) and \ref{wignercat}(f)-\ref{wignercat}(h) we show the Wigner functions $W_{\pm}(\zeta)$ for the generated states (at time $t_{c}$) with various values of the decay rates. Here, the state are obtained by solving quantum master equation~(\ref{masterequation}) with $\hat{H}^{\prime}\rightarrow\hat{H}_{\text{app}}$. The corresponding initial state is given by $|\Phi(0)\rangle =[(|0\rangle_{a}+|1\rangle_{a}) |0\rangle_{b}]/\sqrt{2}$ and a measurement with the bases $|\varphi_{\pm}\rangle_{a}$ is performed at time $t_{c}$ on mode $a$. We can see from Figs.~\ref{wignercat}(d) and \ref{wignercat}(h) that the coherence effect in the cat states becomes weaker for larger decay rates, which means that the decay of the system will attenuate quantum coherence in the generated cat states.

It can be seen from Eq.~(\ref{timestate}) that, at time $t_{s}=(2k+1)\pi/2\varpi_{1}$ for natural numbers $k$, a superposition of the coherent squeezed state and vacuum state can be created with the maximum squeezing strength. If we re-express the state of mode $a$ with the basis states $|\varphi_{\pm}\rangle_{a}$ at time $t_{s}$, then the state of the system becomes
\begin{equation}
|\Psi(t_{s})\rangle =\frac{1}{2}\left[\frac{1}{\mathcal{N}_{+}}|\varphi_{+}\rangle_{a}|\phi_{+}(t_{s})\rangle_{b}+\frac{1}{\mathcal{N}_{-}}%
|\varphi_{-}\rangle_{a}|\phi_{-}(t_{s})\rangle_{b}\right] ,
\end{equation}
where we introduce the superposed states of the coherent squeezed state and vacuum state
\begin{equation}
|\phi_{\pm}(t_{s})\rangle_{b}=\mathcal{N}_{\pm}\{|0\rangle_{b}\pm e^{-i\epsilon(t_{s})}\hat{D}_{b}[\alpha_{1}(t_{s})]\hat{S}_{b}(2\eta
_{1})|0\rangle_{b}\},
\end{equation}
with $\mathcal{N}_{\pm }=\{2\pm 2e^{-\left\vert \alpha _{1}(t_{s})\right\vert^{2}/2} $Re$[e^{-i\epsilon (t_{s})-\alpha _{1}^{\ast 2}(t_{s})\tanh
(2\eta _{1})/2}]/\sqrt{\cosh (2\eta _{1})}\}^{-1/2}$ are the normalization constants.
Mode $b$ will collapse into the states $|\phi_{\pm}(t_{s})\rangle_{b}$ when the states $|\varphi_{\pm}\rangle_{a}$ are measured with measuring probabilities $\mathcal{P}_{\pm}(t_{s})=1/(4\mathcal{N}_{\pm}^{2})$, respectively.
\begin{figure}[tbp]
\center
\includegraphics[width=0.98 \textwidth]{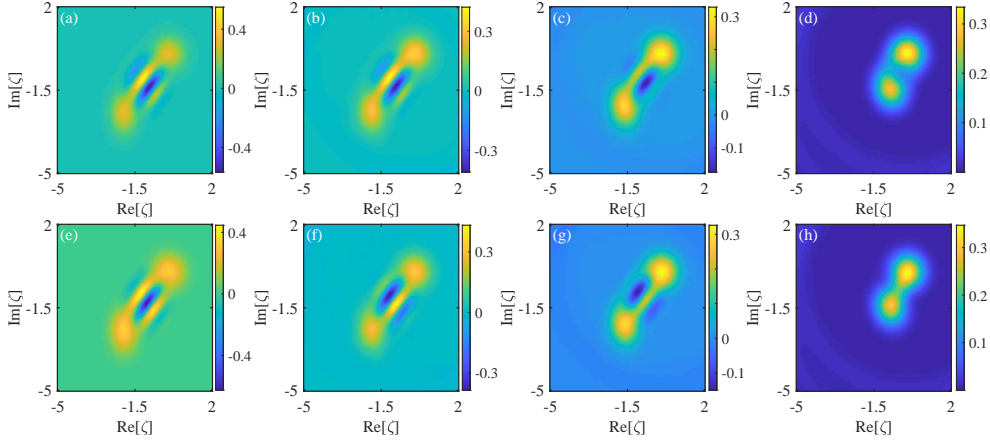}
\caption{The Wigner functions $W_{+}(\zeta)$ ($W_{-}(\zeta)$) of the generated states $\hat{\rho}_{b}^{(+)}(t_{s})$ ($\hat{\rho}_{b}^{(-)}(t_{s})$) are shown in the first (second) row, when the decay rates take various values: (a), (e) $\kappa_{a}/\Delta_{b}=\kappa_{b}/\Delta_{b}=0$; (b), (f) $\kappa_{a}/\Delta_{b}=\kappa_{b}/\Delta_{b}=0.02$; (c), (g) $\kappa_{a}/\Delta_{b}=0.02$ and $\kappa_{b}/\Delta_{b}=0.1$; and (d), (h) $\kappa_{a}/\Delta_{b}=0.02$ and $\kappa_{b}/\Delta_{b}=0.5$. Other parameters used are $t_{s}=\pi/2\varpi_{1}$, $r_{e}=r$, $\theta_{e}=0$, $\bar{n}_{\text{th}}=0$, $\Omega_{p}/\Delta_{b}=0.49$, $\chi/\Delta_{b}=0.01$, and Re$[\varepsilon\exp(i\theta_{d})]/\Delta_{b}=0.38$ corresponding to $g_{1}/\tilde{\omega}_{b}=3.012$ and $g_{2}/\tilde{\omega}_{b}=0.125$.}
\label{csvwigner}
\end{figure}

For the generated states $|\phi_{\pm}(t_{s})\rangle_{b}$, we show the Wigner functions $W_{\pm}(\zeta)$ in Figs.~\ref{csvwigner}(a) and \ref{csvwigner}(e). Here, we also see the clear quantum interference pattern and evidence of macroscopically distinct superposition components. Similar to the state generated at time $t_{c}$, here we also study the influence of the system dissipations on the state generation. In Figs. \ref{csvwigner}(b)-\ref{csvwigner}(d) and \ref{csvwigner}(f)-\ref{csvwigner}(h), we show the Wigner functions $W_{\pm}(\zeta)$ for the generated states (at times $t_{s}$) with various values of the decay rates. The density matrices of the generated states are also obtained by using the same procedure as the state generation case corresponding to the measurement $t_{c}$. However, for the present case, the measurement time is $t_{s}$. Figures \ref{csvwigner}(d) and \ref{csvwigner}(h) indicate that the same conclusion as the previous case. Here, the dissipations will also attenuate the quantum coherence in the generated states.

\section{Discussions on the experimental implementation\label{sec5}}

In this section, we present some discussions about the experimental implementation of this quantum simulation scheme. The physical model proposed in this paper is general and it consists of four main physical elements: (i) the cross-Kerr interaction between the two modes; (ii) the single-excitation driving on mode $b$; (iii) the two-excitation driving on mode $b$; (iv) the squeezed vacuum reservoir of mode $b$. In principle, this scheme can be implemented with physical setups in which the above four physical processes can be realized. Below, we focus our discussions on the superconducting quantum circuits because the above four elements can be well implemented with this setup. In circuit-QED systems, the cross-Kerr interaction has been realized between two superconducting cavities~\cite{Majer2007Nature,Nigg2012PRL,Hoi2013PRL,Holland2015PRL}. In particular, the single-photon cross-Kerr interaction can be resolved from the quantum noise. The single-photon driving process can be realized by driving the cavity field by microwave signal, and the two-photon driving process can be obtained by introducing a two-photon parametric process~\cite{Wang2015PRA,Zhao2018PRA}. In addition, the squeezed vacuum reservoir of the cavity mode can be implemented by injecting a squeezing vacuum field to the cavity.

Below, we present some analyses on the possible experimental parameters. In this paper, we adopt the detuning $\Delta_{b}$ as the frequency scale for convenience in the discussions on the resolved-sideband condition. The value of $\Delta_{b}$ is a controllable value by tuning the driving frequency $\omega_{d}$. For comparison with typical optomechanical systems, we choose $\Delta_{b}\sim2\pi \times 1 - 10$ MHz. In our scheme, the cross-Kerr interaction magnitude could be smaller or larger than the decay rate of the cavity mode $a$. The two-excitation driving $\Omega_{p}$ is smaller than $\Delta_{b}/2$, and the phase $\theta_{p}=\pi$. The single-excitation driving $\varepsilon$ is a free choosing parameter because the displacement amplitude $\alpha_{\text{ss}}$ is proportional to $\varepsilon$. The phase angle $\theta_{d}$ is determined by $\tan\theta_{d}=\kappa_{b}/[2(\Delta_{b}-2\Omega_{p})]$. The parameters $r_{e}=r$ and $\theta_{e}=0$ of the squeezed vacuum reservoir are chosen for suppressing the excitations caused by the two-photon driving, where $r$ is determined by the ratio $\Omega_{p}/\Delta_{b}$. The decay rates of the two modes are taken as $\kappa_{a}/\Delta_{b}=\kappa_{b}/\Delta_{b}\approx0.01-0.1$, which corresponds to $\kappa_{a}$, $\kappa_{b}\sim2\pi \times 10 \ \mathrm{kHz} - 1 \ \mathrm{MHz}$. Based on the above analyses, we know that the present scheme should be within the reach of current or near-future experimental conditions.

\section{Conclusion\label{sec6}}

In conclusion, we have proposed a reliable scheme to implement a quantum simulation for the tunable and ultrastrong mixed-optomechanical interactions. This was realized by introducing both the single- and two-excitation drivings to one of the two bosonic modes coupled via the cross-Kerr interaction. The validity of the approximate Hamiltonians was evaluated by checking the fidelity between the approximate and exact states. The results indicate that the approximate Hamiltonians work well in a wide parameter space. We have also studied the generation of the Schr\"{o}dinger cat states based on the simulated mixed optomechanical interactions. The quantum coherence effect in the generated states has been investigated by calculating the Wigner functions in both the closed- and open-system cases. This work will open a route to the observation of ultrastrong optomechanical effects based on quantum simulators and to the applications of optomechanical technologies in modern quantum sciences.

\begin{backmatter}
\bmsection{Funding}Science and Technology Innovation Program of Hunan Province (2020RC4047); Science and Technology Program of Hunan Province (2017XK2018); National Natural Science Foundation of China (11774087, 11822501, 11935006).

\bmsection{Disclosures}The authors declare no conflicts of interest.

\bmsection{Data availability} Data underlying the results presented in this paper are not publicly available at this time but may be obtained from the authors upon reasonable request.
\end{backmatter}


\end{document}